\documentclass{svproc}
\usepackage{amsmath,bm,bbm,latexsym,soul,nicefrac,young,youngtab}
\usepackage{graphicx,subfigure,multirow,dcolumn}
\usepackage{amssymb}
\usepackage{url}

\def\prd{Physical Review D }

\def\jcap{JCAP }

\def\apj{ApJ }
\def\apjl{ApJL }

\def\aap{A\&A }

\def\araa{ARAA }

\def\mnras{MNRAS }

\def\nar{New Astron. Rev.}

\def\physrep{Phys. Rep.}

\def\sovast{Sov. Astron.}

\def\R{{\cal R}}

\def\lsim{\;\raise0.3ex\hbox{$<$\kern-0.75em\raise-1.1ex\hbox{$\sim$}}\;}
\def\gsim{\;\raise0.3ex\hbox{$>$\kern-0.75em\raise-1.1ex\hbox{$\sim$}}\;}

\def\beq#1{\begin{equation}\label{#1}}
\def\eeq{\end{equation}}
\def\beqa#1{\begin{eqnarray}\label{#1}}
\def\eeqa{\end{eqnarray}}

\def\myfrac#1#2{\left(\frac{#1}{#2}\right)}

\def\comment#1{\relax}

\begin{document}
\mainmatter              % start of a contribution
\title{Primordial Intermediate-mass Binary Black Holes as Targets for Space Laser Interferometers}
\titlerunning{Intermediate-mass black holes}  % abbreviated title (for running head)
\author{Konstantin Postnov\inst{1} \and Ilya Chekh\inst{1}}
\authorrunning{K.Postnov and I.Chekh} % abbreviated author list (for running head)
%Craig Chambers, Kim B. Bruce, and Elisa Bertino}
%
\institute{Sternberg Astronomical Institute, Universitetskij 13, 119234 Moscow, Russia\\
\email{director@sai.msu.ru}}
%,\\ WWW home page:
%\texttt{http://users/\homedir iekeland/web/welcome.html}
%\and
%Universit\'{e} de Paris-Sud,
%Laboratoire d'Analyse Num\'{e}rique, B\^{a}timent 425,\\
%F-91405 Orsay Cedex, France}

\maketitle              % typeset the title of the contribution

\begin{abstract}
%The abstract should summarize the contents of the paper
%using at least 70 and at most 150 words. It will be set in 9-point
%font size and be inset 1.0 cm from the right and left margins.
%There will be two blank lines before and after the Abstract. \dots
% We would like to encourage you to list your keywords within
% the abstract section using the \keywords{...} command.
Primordial black holes (PBHs) with log-normal mass spectrum with masses up to $\sim 10^4-10^5 M_\odot$
can be created after QCD phase transition in the early Universe at $z\sim 10^{12}$ by the modified Affleck-Dine baryogenesis. Using a model binary PBH formation, the expected detection rate of such binary intermediate-mass PBHs by the TianQin space laser interferometer is calculated to be from a few to hundreds events per year for the assumed parameters of the PBH log-normal mass spectrum and abundance  consistent with LIGO-Virgo-KAGRA results. Distinctive features of such primordial IMBH mergings are vanishingly small effective spins, possible high redshifts $z>20$ and lack of association with gas-rich regions or galaxies.
\end{abstract}
\section{Intermediate-mass black holes: a brief introduction}

Black holes (BHs) are topical objects of astrophysical studies. Presently, two broad classes of BHs with different masses are observed by  astronomical methods. Stellar-mass black holes ($\sim 3-60 M_\odot$) are observed in X-ray binaries and found by ground-based gravitational-wave interferometers \cite{2022NewAR..9401642M}. Supermassive BH (SMBHs, $M\sim 10^5-10^9 M_\odot$) are observed in centers of galaxies and active galactic nuclei \cite{2013ARA&A..51..511K}. The intermediate-mass black holes (IMBHs) with masses $\sim 100-10^5 M_\odot$ remain under-detected, likely due to selection effects (see, e.g., \cite{2023arXiv231112118A}). They are possibly found in centers of dwarf galaxies and globular clusters \cite{2024arXiv240600923H,2024arXiv240506015H,2024A&A...684L..19P}. IMBHs can be formed in different ways. They can collapse from very massive Population III stars \cite{2002ApJ...571...30S}, result from dynamical evolution in dense star clusters \cite{2024arXiv240411646G,Fujii_2024} or even have a primordial origin \cite{2016JCAP...11..036B}. IMBHs in galactic centers can underlie tidal disruption events \cite{2024arXiv240410036E}, serve as seeds for SMBH formation \cite{2017IJMPD..2630021M,2024ApJ...965L..21K} and be important sources for space laser gravitational-wave interferometers \cite{2023ApJ...944...81F}. 

In this contribution, we address the possibility to detect primordial binary IMBH mergings by TianQin laser space interferometer \cite{TianQin:2015yph,TianQin:2020hid}. The specific feature of such events could be their occurrence at high redshifts, $z>20$, before early star formation, where no binary IMBHs are expected to appear by conventional astrophysical mechanisms.

\section{Primordial binary black holes}

Primordial black holes formed in the radiation-dominated era in the early Universe were proposed in the seminal papers \cite{1967SvA....10..602Z,1975ApJ...201....1C}. 
%Since then this hypothetical class of astrophysical objects has attracted attention of researchers in various astrophysical aspects. 
Presently, PBH masses can range from $\sim 10^{15}$~g to supermassive BH values of billion solar masses. Their formation, evolution and observational constraints are subjects of intensive studies (see, for example, reviews \cite{2018PhyU...61..115D,2024PhR..1054....1C} and references therein). Binary PBHs became especially interesting after the first LIGO discovery of merging binary stellar-mass black holes   \cite{2016PhRvL.116f1102A}, which may have a primordial origin \cite{2016PhRvL.116t1301B,2016PhRvL.117f1101S,2016JCAP...11..036B}. In many subsequent papers, the published results of LIGO-Virgo-KAGRA collaboration \cite{2021_GWTC3binaries} were satisfactorily modeled by different groups in terms of binary PBHs \cite{2020JCAP...12..017D,2023PhRvD.107f3035L,2024arXiv240505732A}.

PBHs can comprise significant fraction of cold dark matter \cite{1974MNRAS.168..399C}, be seeds for supermassive black holes in galactic centers \cite{2016JCAP...11..036B,2023arXiv231204085L} and globular clusters \cite{2017JCAP...04..036D}, generators of cosmic structure \cite{2018MNRAS.478.3756C}, and are invoked to solve the present-day questions with early galaxy formation suggested by JWST observations with modest requirements on their abundance \cite{2024A&A...685L...8C}.

\subsection{PBHs with extended mass spectrum. \textit{Caveat emptor.}}

The common wisdom is that the PBH can be produced in the early Universe from primordial density perturbations with mass equal to the mass within the cosmological horizon at the time of the PBH formation \cite{1967SvA....10..602Z,1975ApJ...201....1C}: at the radiation-dominated stage, $M_h= m_{pl}^2t$ (in natural units $\hbar=c=k_B=1, G=1/m_{pl}^2$, $m_{pl}$ is the Planck mass)), or $M_h\approx 4.4\times 10^{38}[g] (t/1s)$. The PBH formation from density perturbations is model-dependent (see, e.g., \cite{2024arXiv240513259Y}). Here we will concentrate on the particular PBH fromation model 
with extended log-normal mass spectrum suggested in \cite{1993PhRvD..47.4244D,2009NuPhB.807..229D}:
\beq{e:dndM}
\frac{dn}{dM} = \mu^2 \exp \left[- \gamma \ln^2\left( \frac{M}{M_0} \right) \right] 
\eeq
where $\gamma$ is a dimensionless constant and the parameters $\mu$ and $M_0$ have dimension of mass. In this model, PBHs arise from isocurvature perturbations  with high baryonic charge ('high baryon bubbles', HBBs) created before the end of the inflationary stage. These perturbations turned into large density perturbations at the time of the QCD phase transition, $T_{QCD}\sim 100-150$~MeV, when massless quarks acquire masses from primordial quark-gluon plasma. The expected mean PBH mass is then $M_0\simeq 8M_\odot(100\mathrm{MeV}/T_{QCD})^2$ \cite{2020JCAP...07..063D}. Note that temperature of the QCD phase transition is still uncertain and depends on the chemical potential (matter density), possible magnetic fields, etc. For example, recent holographic calculations \cite{2023EPJC...83...79A} show that $T_{QCD}$ can be appreciably decreased for chemical potential $\mu\sim 0.3$~GeV. Interestingly, this is consistent with expected baryon asymmetry (baryon to photon ratio $\eta=(n_b-n_{\bar b})/n_\gamma$) inside HBBs: to create a PBH with mass of the order of $M\sim M_h$, the density contrast should be $\delta\rho/\rho\ge (M_h/M)^2\sim 1$ at the QCD phase transition, corresponding to $\eta=(\delta\rho/\rho)T_{QCD}/m_q\sim T_{QCD}/m_q\sim 100 \mathrm{MeV}/300 \mathrm{MeV}\sim 0.3$ (here $m_q$ is the mass of constituent quarks). Thus the decreasing $T_{QCD}$ with non-zero chemical potential helps creating PBHs with higher central mass. 

\subsection{Intermediate-mass PBHs}

The extended mass spectrum of PBHs raises the question: how many intermediate-mass PBH can we expect? To substantiate the answer, we will assume the log-normal PBH mass spectrum (\ref{e:dndM}). As was shown in \cite{2016JCAP...11..036B}, the HBBs formed before the end of inflation by the modified Affleck-Dine mechanism  \cite{1993PhRvD..47.4244D} could acquire the baryonic charge up to $\sim 10^{4} M_\odot$ and produce heavy PBHs after the QCD phase transition.  For the log-normal mass distribution (\ref{e:dndM}) the generic mass function $\psi(M)=(M/\rho_{pbh})(dn/d\ln M)$ ($\int \psi(M)d\ln M=1$), where $\rho_{pbh}=f_{pbh}\Omega_{dm}\rho_{cr}$ is the energy density of PBH and $f_{pbh}\le 1$ is the PBH fraction in dark matter, reads:
\beq{e:psi}
\psi(M) =\sqrt{\frac{\gamma}{\pi}}e^{-\frac{1}{\gamma}}\myfrac{M}{M_0}^2e^{-\gamma\ln^2(M/M_0)}\,.
\eeq
Below we shall use masses in units of the central mass $M_0$, $m\equiv M/M_0$. Motivated by LVK results \cite{2020JCAP...12..017D,2023PhRvD.107f3035L,2024arXiv240505732A}, for numerical estimates below we will use $M_0=10,17,25 M_\odot$ and $\gamma=1$. We also will use the standard flat $\Lambda CDM$ cosmological model with $H_0=70$~km/s/Mpc, $\Omega_m=0.3$ and $\Omega_\Lambda=0.7$ where necessary.  
The exponentially small tail of this distribution may seem to imply vanishing fraction of IMBHs. However, integration over the space volumes up to large redshifts available for detection by space interferometers (see \cite{2023ApJ...944...81F} and below) and an almost inverse dependence of the primordial binary IMBH merging rate on the comic time $t(z)$ compensate this small fraction and result in a reasonable probability to detect such mergings by TianQin in several years of observations.

\begin{figure}
\begin{center}
\includegraphics[width=0.8\linewidth]{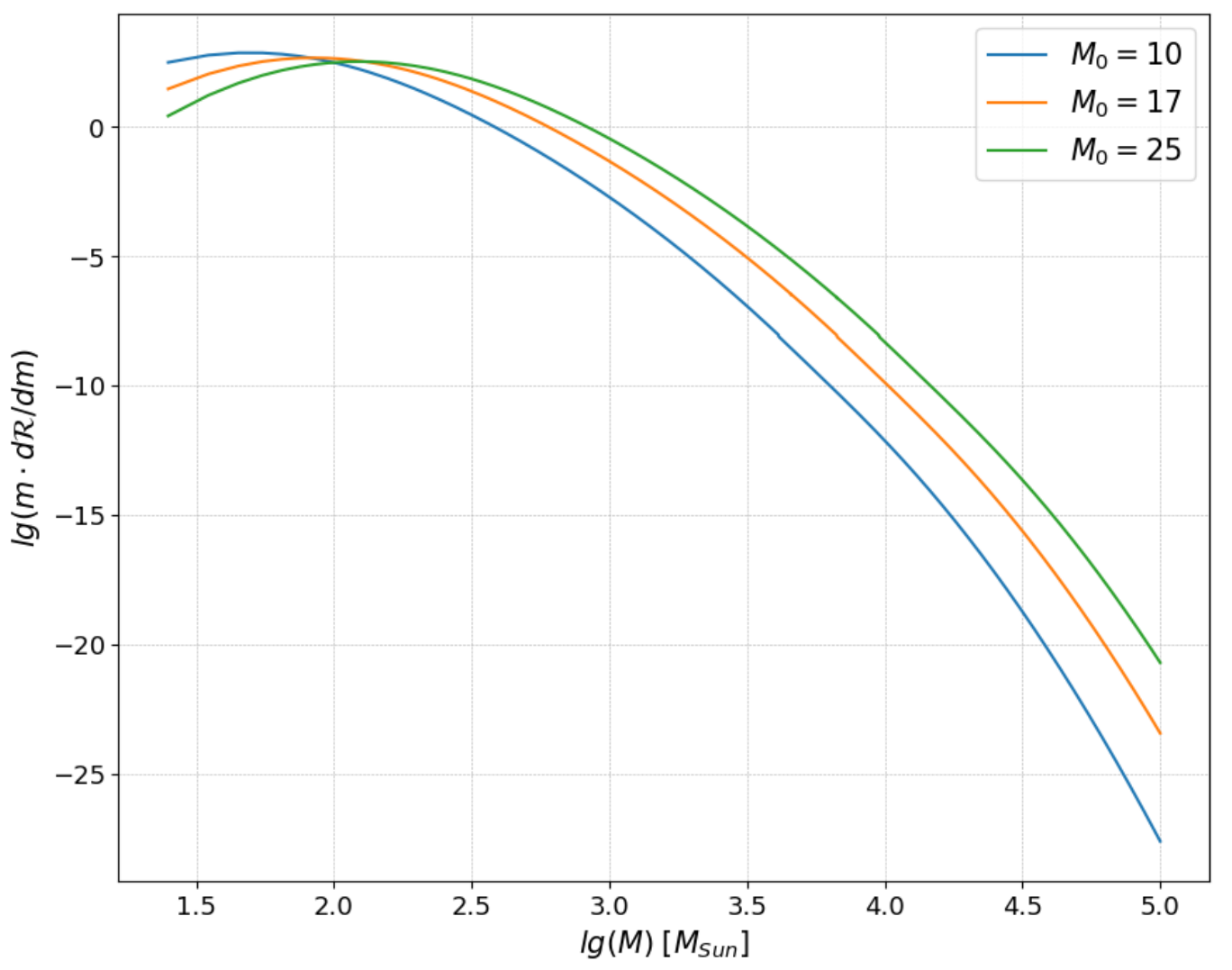}
\caption{Expected distribution of the primordial binary IMBH rate over total mass $md\R/dm$ [Gpc$^{-3}$yr$^{-1}$] at $z=0$ for $f_{pbh}=10^{-3}$, $\gamma=1$ and different $M_0=10\,, 17\,, 25M_\odot$.}
\label{f:mdrdm}
\end{center}
\end{figure}

To make quantitative estimates, we need to specify the PBH binary formation model. Here we will use the model developed in \cite{2019JCAP...02..018R}. This model gives the merging rate of binary PBHs as a function of redshift $z$
\beq{e:dRdm1dm2}
\frac{d\R(z)}{d\ln m_1d\ln m_2}=\frac{1.6\times 10^6} {\mathrm{Gpc}^{3}\mathrm{yr}} S(m_1,m_2,z) f_{pbh}^{\frac{53}{57}}\myfrac{t(z)}{t_0}^{-\frac{34}{37}}\psi(m_1)\psi(m_2)M^{-\frac{32}{37}}\eta^{-\frac{34}{37}}
\eeq
Here $S\le 1$ is the suppression factor \cite{2019JCAP...02..018R,2021JCAP...03..068H} which we will set to 1 below (justified by $f_{pbh}\ll 1$), $t_0$ is the age of the Universe, $M=(m_1+m_2)/M_\odot$, $\eta=m_1m_2/(m_1+m_2)^2$. It is convenient to change variables $m_1,m_2 \to m,q$, where $m=m_1+m_2$ is the total binary mass in units $M_0$ and $q=m_2/m_1\le 1$ is the binary mass ratio.  Substituting $m_1=m/(1+q)$, $m_2=mq/(1+q)$ and the Jacobian $J(m,q)=m/(1+q)^2$ into (\ref{e:dRdm1dm2}) we get 
\begin{eqnarray}
\label{e:dRdmdq}
\frac{d\R(z)}{dmdq}&=\underbrace{\frac{1.6\times 10^6} {\mathrm{Gpc}^{3}\mathrm{yr}}f_{pbh}^{\frac{53}{57}}\myfrac{t(z)}{t_0}^{-\frac{34}{37}}\myfrac{\gamma}{\pi}e^{-\frac{2}{\gamma}}\myfrac{M_\odot}{M_0}^{-\frac{32}{37}}}_A\nonumber \\
&\times \frac{m^2q}{(1+q)^2}e^{-\gamma\ln^2(\frac{m}{1+q})}e^{-\gamma\ln^2(\frac{mq}{1+q})}m^{-\frac{32}{37}}\myfrac{q}{(1+q)^2}^{-\frac{34}{37}}\frac{m}{(1+q)^2}\nonumber\\
& = A\times m^{(3-\frac{32}{37})}e^{-2\gamma\ln^2m} \nonumber \\
&\times\underbrace{
%{\myfrac{q}{(1+q)^2}^{-2\gamma\ln m} 
e^{-\gamma(\ln^2\frac{1}{1+q}+\ln^2\frac{q}{1+q})}\myfrac{q}{(1+q)^2}^{1-\frac{34}{37}-2\gamma\ln m}\frac{1}{(1+q)^2}}_{\Phi(q,m)}.
\end{eqnarray}
Performing integration over $q$ we obtain the specific rate
\beq{e:drdM}
\frac{d\R(z)}{dm}=A\times m^{(3-\frac{32}{37})}e^{-2\gamma\ln^2m}\int\limits_0^1 \Phi(q,m) dq\,.
\eeq
The specific binary PBH coalescence rate $md\R/dm$ at $z=0$ for several fiducial model parameters is shown in Fig. \ref{f:mdrdm}. Note that by adjusting $f_{pbh}$,  at low masses ($M\sim 100 M_\odot$) this rate can be easily made consistent with the observed binary BH merging rate ($\sim 100 \,\mathrm{Gpc}^{-3}\mathrm{yr}^{-1}$ at $z\sim 0$) as inferred from LIGO-Virgo-KAGRA observations \cite{2021_GWTC3binaries}. 
\begin{figure}
\begin{center}
\includegraphics[width=0.8\linewidth]{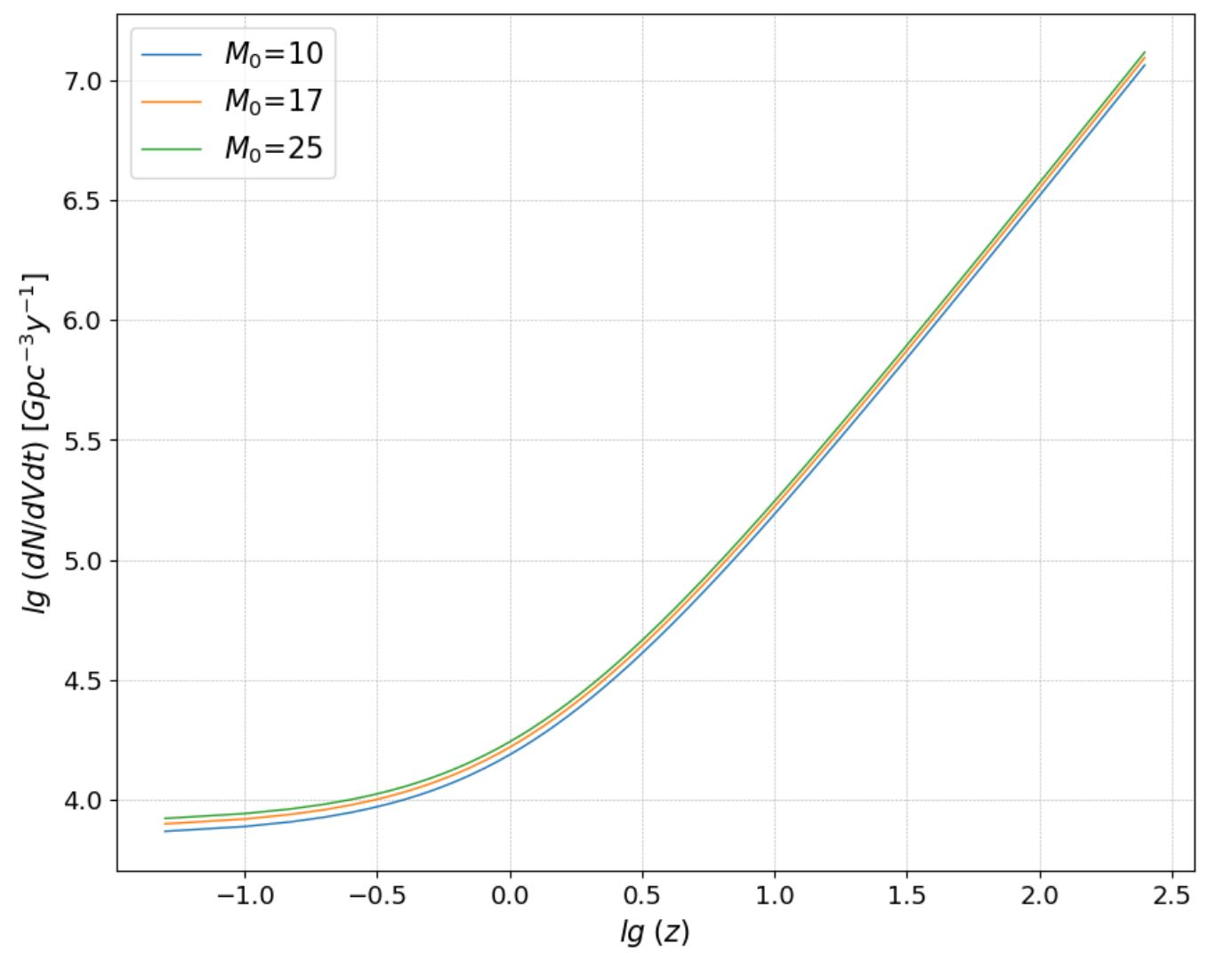}
\caption{The comoving merging rate of primordial binary BH (\ref{e:dRdm1dm2}) as a function of redhift $\R(z)\sim t(z)^{-34/37}$.}
\label{f:R(z)}
\end{center}
\end{figure}
The integration of $d\R(z)/dm$ over $m$ yields the total binary PBH merging rate as a function of redshift (Fig. \ref{f:R(z)}). Unlike astrophysical binary BH, in the adopted model the binary PBH merging rate monotonically increases with redshift as 
$\R(z)\sim 1/t(z)$ at all $z$.
We note that the calculated here specific binary IMBH merging rate does not contradict the SMBH merging rate inferred from the recent pulsar timing measurements 
(see, e.g., \cite{2024A&A...685A..94E}).

\section{Prospects for detection of merging binary primordial IMBH by TianQin}

\begin{figure}
\begin{center}
\includegraphics[width=0.8\linewidth]{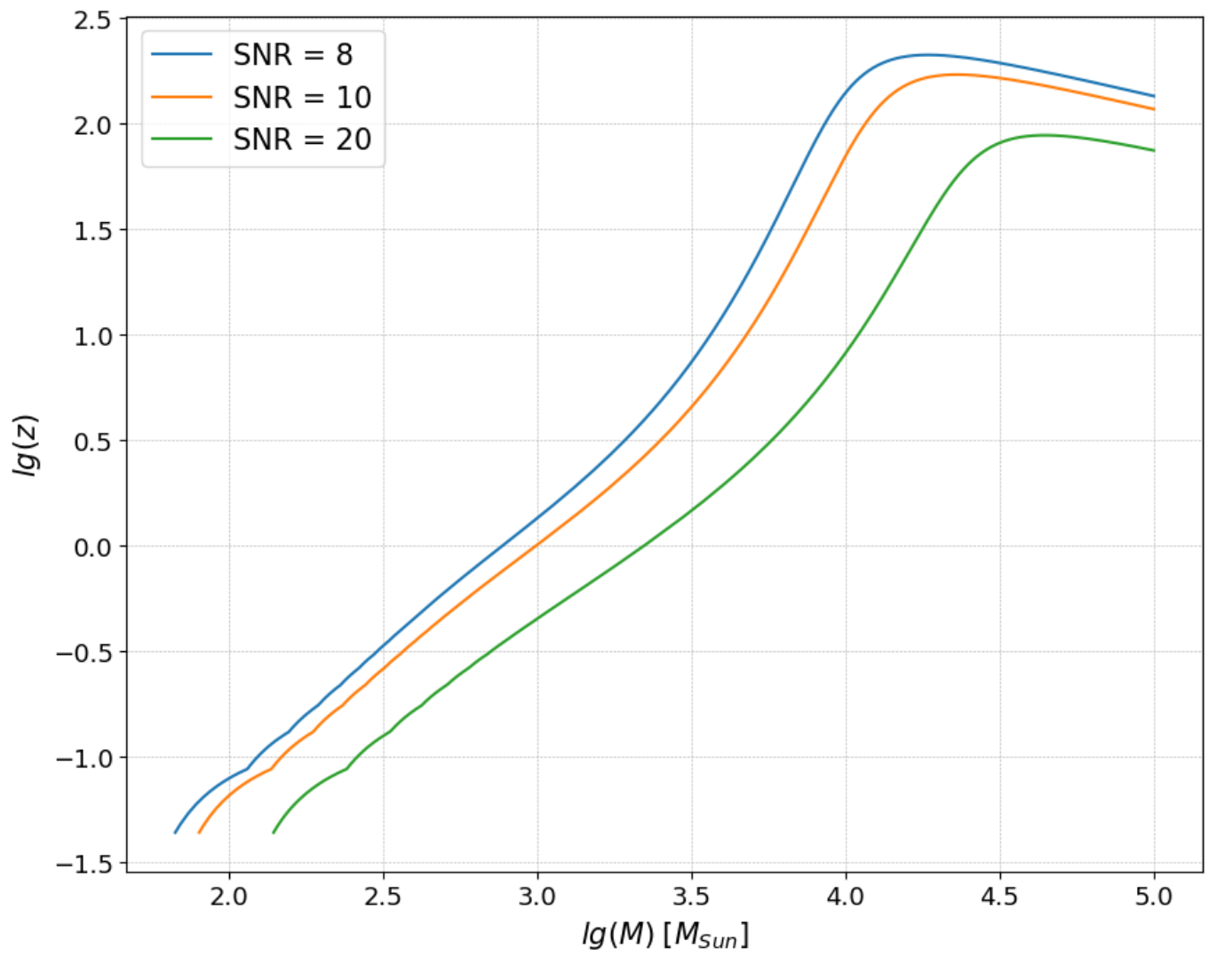}
\caption{Limiting redshift $z_{lim}(m)$  for detection of coalescing binary BHs with total mass $m$ (in the source's frame) and mass ratio  $q=0.5$ at TianQin threshold SNR=8, 10, 20. }
\label{f:zlim}
\end{center}
\end{figure}

The detector sensitivity noise curve imposes a limiting distance (detection horizon) from which a binary with total mass $m$ and mass ratio $q$\footnote{For binary coalescences, SNR$\propto {\cal M}^{5/6}$, where ${\cal M}=M\frac{q^{3/5}}{(1+q)^{6/5}}$ is the binary chirp mass, is a function of the binary total mass $M$ and mass ratio $q$.} can be registered with a given signal-to-noise ratio (SNR). We take the TianQin sensitivity and perform the SNR calculation after \cite{2019PhRvD..99l3002F}. The limiting redshit $z_{lim}(m)$ for different SNR and $q=0.5$ is shown in Fig. \ref{f:zlim}.

\begin{figure}
\begin{center}
\includegraphics[width=0.8\textwidth]{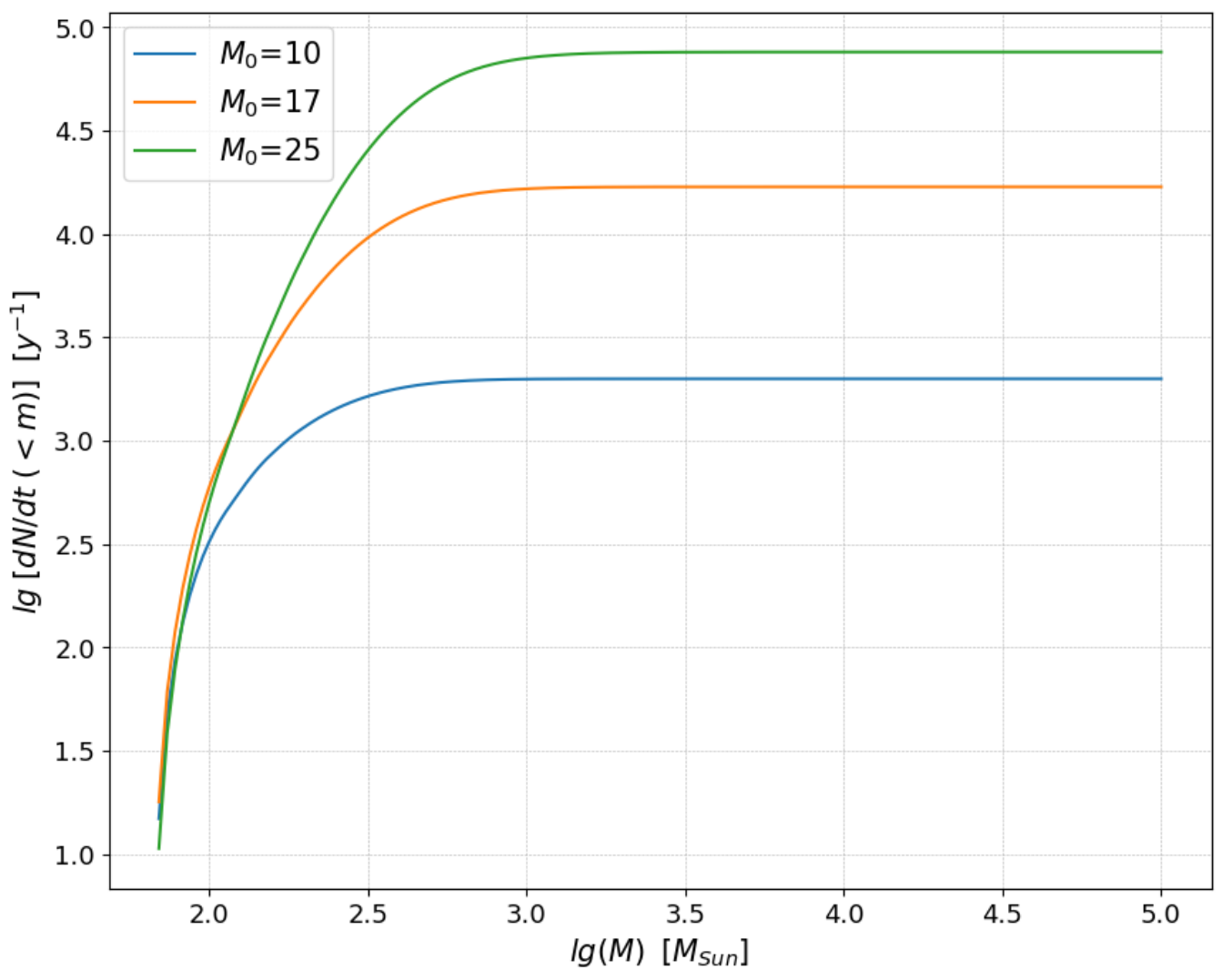}
\caption{Expected cumulative detection rate $\frac{dN}{dt}(\le m)$ [yr$^{-1}$]  of the primordial binary IMBH rate with total mass $m$  by TianQin interferometer for for SNR=8,  $f_{pbh}=10^{-3}$, $\gamma=1$.}
\label{f:FRa}
\end{center}
\end{figure}

\begin{figure}
\begin{center}
\includegraphics[width=0.8\textwidth]{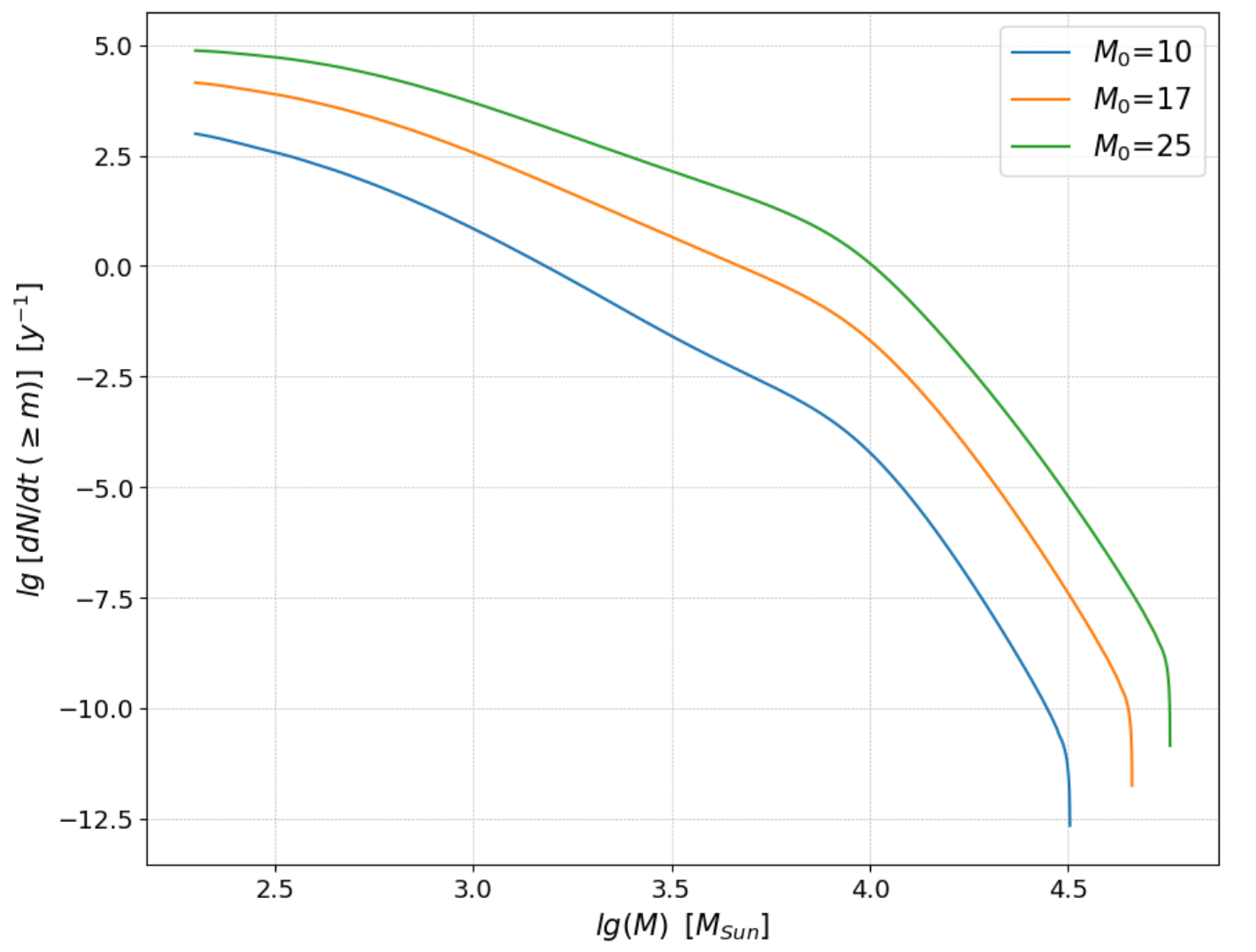}
\caption{The same as Fig.4 for $\frac{dN}{dt}(\ge m)$. }
\label{f:FRb}
\end{center}
\end{figure}

The expected merging rate of binary PBH with taking into account the detector sensitivity is obtained by integrating the specific rate $d\R(z)/dm$ over redshift up to $z_{lim}(m)$:
\beq{e:Rdet}
\frac{dN}{dtdm}[\mathrm{yr}^{-1}]=\int\limits_0^{z_{lim}(m)} \frac{d\R(z)}{dm}\frac{1}{(1+z)}\frac{dV(z)}{dz}dz\,,\quad \frac{dN}{dt}(>m)=\int\limits_m^\infty \frac{dN}{dtdm} dm
\eeq
The obtained cumulative detection rate $\frac{dN}{dt}(\le m)$ and $\frac{dN}{dt}(\ge m)$ are shown in  Fig. \ref{f:FRa} and Fig. \ref{f:FRb}, respectively. It is seen that with the planned TianQin sensitivity and the adopted assumptions on primordial binary IMBH formation, the detection of a few mergings of primordial binary IMBHs is possible. A unique signature of such mergings would be their zero effective spins and high redshift $z>20$ where no other but primordial black holes are thought to exist.

\section{Conclusions}

IMBHs with masses from several hundred to ten thousands solar masses are very topical in modern astrophysics and cosmology. Their formation and subsequent evolution (accretion mass growth in galactic centers, coalescence rate of binary IMBHs) remain disputable. Here we have focused on a particular class  of IMBHs, primordial IMBHs. These objects can emerge at the radiation-dominated stage in the early  Universe during the QCD phase transition from primordial isocurvature  fluctuations with large baryonic number (HBBs) with the baryon asymmetry parameter $\eta\lesssim 1$, which can be prepared before the end of inflation by a complex scalar field with baryonic charge coupled to inflaton in the modified  Affleck-Dine baryogenesis proposed in \cite{1993PhRvD..47.4244D}. This mechanism predicts the existence of log-normal distribution of HBB baryonic mass they acquire after QCD phase transition at $T_{QCD}\sim 100$~MeV. The expected non-zero chemical potential inside HBBs $\mu\sim 0.3$~GeV decreases $T_{QCD}$.  Large density fluctuations emerged from HBBs after the QCD phase transition can give rise to primordial black holes with the characteristic mass of the order of the cosmological horizon at that time of about $10-20 M_\odot$ and log-normal mass distribution extending up to $\sim 10^4 M_\odot$ \cite{2016JCAP...11..036B}. 

Using a particular model of  primordial binary BH formation \cite{2019JCAP...02..018R}, we calculated the expected detection rate by such binary PBHs by TianQin interferometer (Fig. \ref{f:FRa},\ref{f:FRb}). It  is shown that for PBH formation parameters consistent with the detection  rate and distribution of binary BH mergings by the LIGO-Virgo-KAGRA interferometers ($M_0\sim 10-20 M_\odot$, $\gamma\sim 1$) and modest PBH fraction in dark matter density $f_{pbh}=10^{-3}$ \cite{2020JCAP...12..017D,2021JCAP...03..068H,2023arXiv231204085L,2024arXiv240505732A}, the expected number of detections of merging IMBHs with total mass $M\sim 10^3-10^4 M_\odot$ can range from a few to hundred events in several years of observations depending on parameters. We conclude that 
space laser gravitational-wave interferometers sensitive at mHz like TinQin will be able to prove the existence of primordial IMBHs having the unique feature of being only binary IMBHs mergings at high redshifts $z>20$ in the dark age epoch. Even at lower redshifts in the stellar epoch, the mergings of primordial IMBHs can not be associated with gas-rich regions or galaxies and thus are not expected to be associated with strong electromagnetic signals. 

\vskip\baselineskip
\textit{Acknowledgments.} We thank A.D. Dolgov for discussions. The authors acknowledge the support from the Russian Science Foundation grant 23-42-00055. KP also thanks the TianQin Research Center for Gravitational Physics in Zhuhai for hospitality.
%\bibliographystyle{acm}
%\bibliography{ref}

\begin{thebibliography}{10}

\bibitem{2016PhRvL.116f1102A}
{\sc {Abbott}, B.~P., {Abbott}, R., {Abbott}, T.~D., {Abernathy}, M.~R., {Acernese}, F., {Ackley}, K., {Adams}, C., {Adams}, T., {Addesso}, P., {Adhikari}, R.~X., and et~al.}
\newblock {Observation of Gravitational Waves from a Binary Black Hole Merger}.
\newblock {\em Physical Review Letters 116}, 6 (Feb. 2016), 061102.

\bibitem{2024arXiv240505732A}
{\sc {Andr{\'e}s-Carcasona}, M., {Iovino}, A.~J., {Vaskonen}, V., {Veerm{\"a}e}, H., {Mart{\'\i}nez}, M., {Pujol{\`a}s}, O., and {Mir}, L.~M.}
\newblock {Constraints on primordial black holes from LIGO-Virgo-KAGRA O3 events}.
\newblock {\em arXiv e-prints\/} (May 2024), arXiv:2405.05732.

\bibitem{2023EPJC...83...79A}
{\sc {Aref'eva}, I.~Y., {Ermakov}, A., {Rannu}, K., and {Slepov}, P.}
\newblock {Holographic model for light quarks in anisotropic hot dense QGP with external magnetic field}.
\newblock {\em European Physical Journal C 83}, 1 (Jan. 2023), 79.

\bibitem{2023arXiv231112118A}
{\sc {Askar}, A., {Baldassare}, V.~F., and {Mezcua}, M.}
\newblock {Intermediate-Mass Black Holes in Star Clusters and Dwarf Galaxies}.
\newblock {\em arXiv e-prints\/} (Nov. 2023), arXiv:2311.12118.

\bibitem{2016PhRvL.116t1301B}
{\sc {Bird}, S., {Cholis}, I., {Mu{\~n}oz}, J.~B., {Ali-Ha{\"i}moud}, Y., {Kamionkowski}, M., {Kovetz}, E.~D., {Raccanelli}, A., and {Riess}, A.~G.}
\newblock {Did LIGO Detect Dark Matter?}
\newblock {\em Physical Review Letters 116}, 20 (May 2016), 201301.

\bibitem{2016JCAP...11..036B}
{\sc {Blinnikov}, S., {Dolgov}, A., {Porayko}, N.~K., and {Postnov}, K.}
\newblock {Solving puzzles of GW150914 by primordial black holes}.
\newblock {\em \jcap 2016}, 11 (Nov. 2016), 036.

\bibitem{2018MNRAS.478.3756C}
{\sc {Carr}, B., and {Silk}, J.}
\newblock {Primordial black holes as generators of cosmic structures}.
\newblock {\em \mnras 478}, 3 (Aug. 2018), 3756--3775.

\bibitem{1975ApJ...201....1C}
{\sc {Carr}, B.~J.}
\newblock {The primordial black hole mass spectrum.}
\newblock {\em \apj 201\/} (Oct. 1975), 1--19.

\bibitem{2024PhR..1054....1C}
{\sc {Carr}, B.~J., {Clesse}, S., {Garc{\'\i}a-Bellido}, J., {Hawkins}, M.~R.~S., and {K{\"u}hnel}, F.}
\newblock {Observational evidence for primordial black holes: A positivist perspective}.
\newblock {\em \physrep 1054\/} (Feb. 2024), 1--68.

\bibitem{1974MNRAS.168..399C}
{\sc {Carr}, B.~J., and {Hawking}, S.~W.}
\newblock {Black holes in the early Universe}.
\newblock {\em \mnras 168\/} (Aug. 1974), 399--416.

\bibitem{2024A&A...685L...8C}
{\sc {Colazo}, P.~E., {Stasyszyn}, F., and {Padilla}, N.}
\newblock {Structure formation with primordial black holes to alleviate early star formation tension revealed by JWST}.
\newblock {\em \aap 685\/} (May 2024), L8.

\bibitem{2017JCAP...04..036D}
{\sc {Dolgov}, A., and {Postnov}, K.}
\newblock {Globular cluster seeding by primordial black hole population}.
\newblock {\em \jcap 2017}, 4 (Apr. 2017), 036.

\bibitem{2020JCAP...07..063D}
{\sc {Dolgov}, A., and {Postnov}, K.}
\newblock {Why the mean mass of primordial black hole distribution is close to 10M$_{solar}$}.
\newblock {\em \jcap 2020}, 7 (July 2020), 063.

\bibitem{1993PhRvD..47.4244D}
{\sc {Dolgov}, A., and {Silk}, J.}
\newblock {Baryon isocurvature fluctuations at small scales and baryonic dark matter}.
\newblock {\em \prd 47}, 10 (May 1993), 4244--4255.

\bibitem{2018PhyU...61..115D}
{\sc {Dolgov}, A.~D.}
\newblock {Massive and supermassive black holes in the contemporary and early Universe and problems in cosmology and astrophysics}.
\newblock {\em Physics Uspekhi 61}, 2 (Feb. 2018), 115.

\bibitem{2009NuPhB.807..229D}
{\sc {Dolgov}, A.~D., {Kawasaki}, M., and {Kevlishvili}, N.}
\newblock {Inhomogeneous baryogenesis, cosmic antimatter, and dark matter}.
\newblock {\em Nuclear Physics B 807}, 1-2 (Jan. 2009), 229--250.

\bibitem{2020JCAP...12..017D}
{\sc {Dolgov}, A.~D., {Kuranov}, A.~G., {Mitichkin}, N.~A., {Porey}, S., {Postnov}, K.~A., {Sazhina}, O.~S., and {Simkin}, I.~V.}
\newblock {On mass distribution of coalescing black holes}.
\newblock {\em \jcap 2020}, 12 (Dec. 2020), 017.

\bibitem{2024arXiv240410036E}
{\sc {Eftekhari}, T., {Tchekhovskoy}, A., {Alexander}, K.~D., {Berger}, E., {Chornock}, R., {Laskar}, T., {Margutti}, R., {Yao}, Y., {Cendes}, Y., {Gomez}, S., {Hajela}, A., and {Pasham}, D.~R.}
\newblock {Late-time X-ray Observations of the Jetted Tidal Disruption Event AT2022cmc: The Relativistic Jet Shuts Off}.
\newblock {\em arXiv e-prints\/} (Apr. 2024), arXiv:2404.10036.

\bibitem{2024A&A...685A..94E}
{\sc {EPTA Collaboration}, {InPTA Collaboration}, {Antoniadis}, J., {Arumugam}, P., {Arumugam}, S., and {et.al.}}
\newblock {The second data release from the European Pulsar Timing Array. IV. Implications for massive black holes, dark matter, and the early Universe}.
\newblock {\em \aap 685\/} (May 2024), A94.

\bibitem{2019PhRvD..99l3002F}
{\sc {Feng}, W.-F., {Wang}, H.-T., {Hu}, X.-C., {Hu}, Y.-M., and {Wang}, Y.}
\newblock {Preliminary study on parameter estimation accuracy of supermassive black hole binary inspirals for TianQin}.
\newblock {\em \prd 99}, 12 (June 2019), 123002.

\bibitem{2023ApJ...944...81F}
{\sc {Fragione}, G., and {Loeb}, A.}
\newblock {Constraining the Cosmic Merger History of Intermediate-mass Black Holes with Gravitational Wave Detectors}.
\newblock {\em \apj 944}, 1 (Feb. 2023), 81.

\bibitem{Fujii_2024}
{\sc Fujii, M.~S., Wang, L., Tanikawa, A., Hirai, Y., and Saitoh, T.~R.}
\newblock Simulations predict intermediate-mass black hole formation in globular clusters.
\newblock {\em Science\/} (May 2024).

\bibitem{2024arXiv240411646G}
{\sc {Gonz{\'a}lez Prieto}, E., {Weatherford}, N.~C., {Fragione}, G., {Kremer}, K., and {Rasio}, F.~A.}
\newblock {IMBH Progenitors from Stellar Collisions in Dense Star Clusters}.
\newblock {\em arXiv e-prints\/} (Apr. 2024), arXiv:2404.11646.

\bibitem{2024arXiv240506015H}
{\sc {H{\"a}berle}, M., {Neumayer}, N., {Seth}, A., {Bellini}, A., {Libralato}, M., {Baumgardt}, H., {Whitaker}, M., {Dumont}, A., {Alfaro Cuello}, M., {Anderson}, J., {Clontz}, C., {Kacharov}, N., {Kamann}, S., {Feldmeier-Krause}, A., {Milone}, A., {Nitschai}, M.~S., {Pechetti}, R., and {van de Ven}, G.}
\newblock {Fast-moving stars around an intermediate-mass black hole in Omega Centauri}.
\newblock {\em arXiv e-prints\/} (May 2024), arXiv:2405.06015.

\bibitem{2024arXiv240600923H}
{\sc {Huang}, Y., {Li}, Q., {Liu}, J., {Dong}, X., {Zhang}, H., {Lu}, Y., and {Du}, C.}
\newblock {A high-velocity star recently ejected by an intermediate-mass black hole in M15}.
\newblock {\em arXiv e-prints\/} (June 2024), arXiv:2406.00923.

\bibitem{2021JCAP...03..068H}
{\sc {H{\"u}tsi}, G., {Raidal}, M., {Vaskonen}, V., and {Veerm{\"a}e}, H.}
\newblock {Two populations of LIGO-Virgo black holes}.
\newblock {\em \jcap 2021}, 3 (Mar. 2021), 068.

\bibitem{2013ARA&A..51..511K}
{\sc {Kormendy}, J., and {Ho}, L.~C.}
\newblock {Coevolution (Or Not) of Supermassive Black Holes and Host Galaxies}.
\newblock {\em \araa 51}, 1 (Aug. 2013), 511--653.

\bibitem{2024ApJ...965L..21K}
{\sc {Kov{\'a}cs}, O.~E., {Bogd{\'a}n}, {\'A}., {Natarajan}, P., {Werner}, N., {Azadi}, M., {Volonteri}, M., {Tremblay}, G.~R., {Chadayammuri}, U., {Forman}, W.~R., {Jones}, C., and {Kraft}, R.~P.}
\newblock {A Candidate Supermassive Black Hole in a Gravitationally Lensed Galaxy at Z {\ensuremath{\approx}} 10}.
\newblock {\em \apjl 965}, 2 (Apr. 2024), L21.

\bibitem{2023arXiv231204085L}
{\sc {Liu}, B., and {Bromm}, V.}
\newblock {Impact of primordial black holes on the formation of the first stars and galaxies}.
\newblock {\em arXiv e-prints\/} (Dec. 2023), arXiv:2312.04085.

\bibitem{2023PhRvD.107f3035L}
{\sc {Liu}, L., {You}, Z.-Q., {Wu}, Y., and {Chen}, Z.-C.}
\newblock {Constraining the merger history of primordial-black-hole binaries from GWTC-3}.
\newblock {\em \prd 107}, 6 (Mar. 2023), 063035.

\bibitem{TianQin:2015yph}
{\sc Luo, J., et~al.}
\newblock {TianQin: a space-borne gravitational wave detector}.
\newblock {\em Class. Quant. Grav. 33}, 3 (2016), 035010.

\bibitem{TianQin:2020hid}
{\sc Mei, J., et~al.}
\newblock {The TianQin project: current progress on science and technology}.
\newblock {\em PTEP 2021}, 5 (2021), 05A107.

\bibitem{2017IJMPD..2630021M}
{\sc {Mezcua}, M.}
\newblock {Observational evidence for intermediate-mass black holes}.
\newblock {\em International Journal of Modern Physics D 26}, 11 (Jan. 2017), 1730021.

\bibitem{2022NewAR..9401642M}
{\sc {Mirabel}, I.~F., and {Rodr{\'\i}guez}, L.~F.}
\newblock {Black holes at cosmic dawn in the redshifted 21cm signal of HI}.
\newblock {\em \nar 94\/} (June 2022), 101642.

\bibitem{2024A&A...684L..19P}
{\sc {Pascale}, R., {Nipoti}, C., {Calura}, F., and {Della Croce}, A.}
\newblock {The central black hole in the dwarf spheroidal galaxy Leo I: Not supermassive, at most an intermediate-mass candidate}.
\newblock {\em \aap 684\/} (Apr. 2024), L19.

\bibitem{2019JCAP...02..018R}
{\sc {Raidal}, M., {Spethmann}, C., {Vaskonen}, V., and {Veerm{\"a}e}, H.}
\newblock {Formation and evolution of primordial black hole binaries in the early universe}.
\newblock {\em \jcap 2019}, 2 (Feb. 2019), 018.

\bibitem{2016PhRvL.117f1101S}
{\sc {Sasaki}, M., {Suyama}, T., {Tanaka}, T., and {Yokoyama}, S.}
\newblock {Primordial Black Hole Scenario for the Gravitational-Wave Event GW150914}.
\newblock {\em Physical Review Letters 117}, 6 (Aug. 2016), 061101.

\bibitem{2002ApJ...571...30S}
{\sc {Schneider}, R., {Ferrara}, A., {Natarajan}, P., and {Omukai}, K.}
\newblock {First Stars, Very Massive Black Holes, and Metals}.
\newblock {\em \apj 571}, 1 (May 2002), 30--39.

\bibitem{2021_GWTC3binaries}
{\sc {The LIGO Scientific Collaboration}}.
\newblock Population of merging compact binaries inferred using gravitational waves through gwtc-3.
\newblock {\em Phys. Rev. X 13\/} (Mar 2023), 011048.

\bibitem{2024arXiv240513259Y}
{\sc {Young}, S.}
\newblock {Computing the abundance of primordial black holes}.
\newblock {\em arXiv e-prints\/} (May 2024), arXiv:2405.13259.

\bibitem{1967SvA....10..602Z}
{\sc {Zel'dovich}, Y.~B., and {Novikov}, I.~D.}
\newblock {The Hypothesis of Cores Retarded during Expansion and the Hot Cosmological Model}.
\newblock {\em \sovast 10\/} (Feb. 1967), 602.

\end{thebibliography}
%%%%%%%%%%%%%%%%%%%%%%%%%%%%%%%%%%%%%%%%%%%%%%%%%%%%%%%%%%%%%%%%

\end{document}